\documentclass[sigconf]{acmart}
\usepackage{multirow}
\renewcommand\footnotetextcopyrightpermission[1]{} 

\usepackage{footnote}
\makesavenoteenv{tabular}
\makesavenoteenv{table}

\usepackage{setspace}
\usepackage{graphicx}
\usepackage{url}

\usepackage{listings}
\usepackage{color}

\definecolor{dkgreen}{rgb}{0,0.6,0}
\definecolor{gray}{rgb}{0.5,0.5,0.5}
\definecolor{mauve}{rgb}{0.58,0,0.82}

\begin{document}

\title{Software Quality Assessment for Robot Operating System}
\vspace{5cm}
\author{Mohannad Alhanahnah}
\affiliation{\institution{University of Nebraska-Lincoln}}
\email{mohannad@huskers.unl.edu}

\begin{abstract}
  Robot Operating System (ROS) is widely used in academia and industry, and importantly is leveraged in safety-critical robotic systems. The quality of ROS software can affect the safety and security properties of robotics systems; therefore, reliability and quality are imperative to guarantee. Source code static analysis is a key approach to formally perform software verification. We address two concerns in this paper: (1) conducting a systematic literature review study to provide a complete picture of the existing methods that analyze different aspects of ROS software, (2) performing empirical study to evaluate software properties that can influence the functionality of ROS. We leverage PMD~\footnote{PMD is not an acronym}, off-the-shelf static analysis tool, to conduct our empirical study over a set of ROS repositories implemented using Java. The survey analysis shows a significant shortcoming in the body of research by the lack of tailored analysis mechanisms for assessing ROS2 code and reveals that the majority of research efforts are centered around ROS1. Our empirical study shows that the Java code of ROS2 does not suffer from serious issues and the majority of the detected alerts are code style issues. 
\end{abstract}

\keywords{static analysis, ROS, software quality, robotics}

\maketitle

\section{Introduction}
    ROS is a framework that provides a communication layer between robotic systems, as well as libraries and tools for the development of robotic systems software. This makes ROS widely adopted in developing robotics systems, which are increasingly used in safety-critical contexts, including health and transportation. However, developers of robotics systems lack applying systematic approaches to develop safety-critical software applying~\cite{Ingibergsson2015}. Especially, the ROS framework is complex and consists of several components that are developed using various programming languages. Therefore, evaluating the software features of this framework is vital and challenging. Several code features are considered in the body of research. Phriky-Units~\cite{Ore2017,Ore2017FP} and Phys~\cite{Kate2018} are static analysis tools investigate the inconsistency of physical units in robotics and cyber-physical systems. Both tools have evaluated the implementation of ROS1. While HAROS~\cite{Santos2016} collects code metrics for several ROS repositories, through a collection of static analysis tools. Other works~\cite{McClean2013,DeMarinis2019} take dynamic analysis approaches to verify the security of ROS. All prior work focuses on analyzing ROS1, but none of them consider ROS2. Moreover, these works do not provide a holistic picture of various programming properties of the ROS framework, because their scope is limited to investigate a specific property (i.e. security, physical units). The investigated ROS projects are centered around ROS implementation in C++, but ROS also provides programming interfaces using other languages such as Java, which has not been considered in prior work.
    
    To address these concerns, First, we conduct a small-scale systematic literature review to understand the scope and limitations of prior work. Second, we perform an empirical study to check the conformance of ROS2 Java projects with coding standards. We leverage an off-the-shelf static analysis tool, is called PMD, which facilitates the analysis of various coding standards, including security, performance and code style. These coding standards are widely enforced in safety-critical contexts to improve software reliability.

    The rest of this paper is organized as follows: Section~\ref{sec:background} highlights the software architecture of ROS2. The applied approach in this work is discussed in Section~\ref{sec:overallApproach}. Section~\ref{sec:approachExp} describes our empirical study and introduces the static analysis tool that is used to perform the empirical study. We present and discuss our results in Section~\ref{sec:study}. Finally, Section~\ref{sec:conclusion} concludes this work and provides potential research directions.

\section{Background}\label{sec:background}
ROS2 is a collection of tools and libraries that aid robotics software development. In this section, we highlight the architecture of this framework, because understanding the architecture of ROS2 is an intrinsic factor to identify the involved programming languages and components to conduct an appropriate evaluation. 

ROS2 implements Object Management Group’s Data Distribution Service (DDS) as the communication middleware~\cite{ers2019ros2}, which enables ROS2 nodes to automatically find each other on the network, thus there is no need for a ROS2 master. Figure~\ref{fig:rosArch} depicts the internal API architecture of ROS2. The architecture consists mainly of two layers~\cite{ROSArch}:

\begin{figure}[h!]
    \centering
    \includegraphics[width=1\columnwidth]{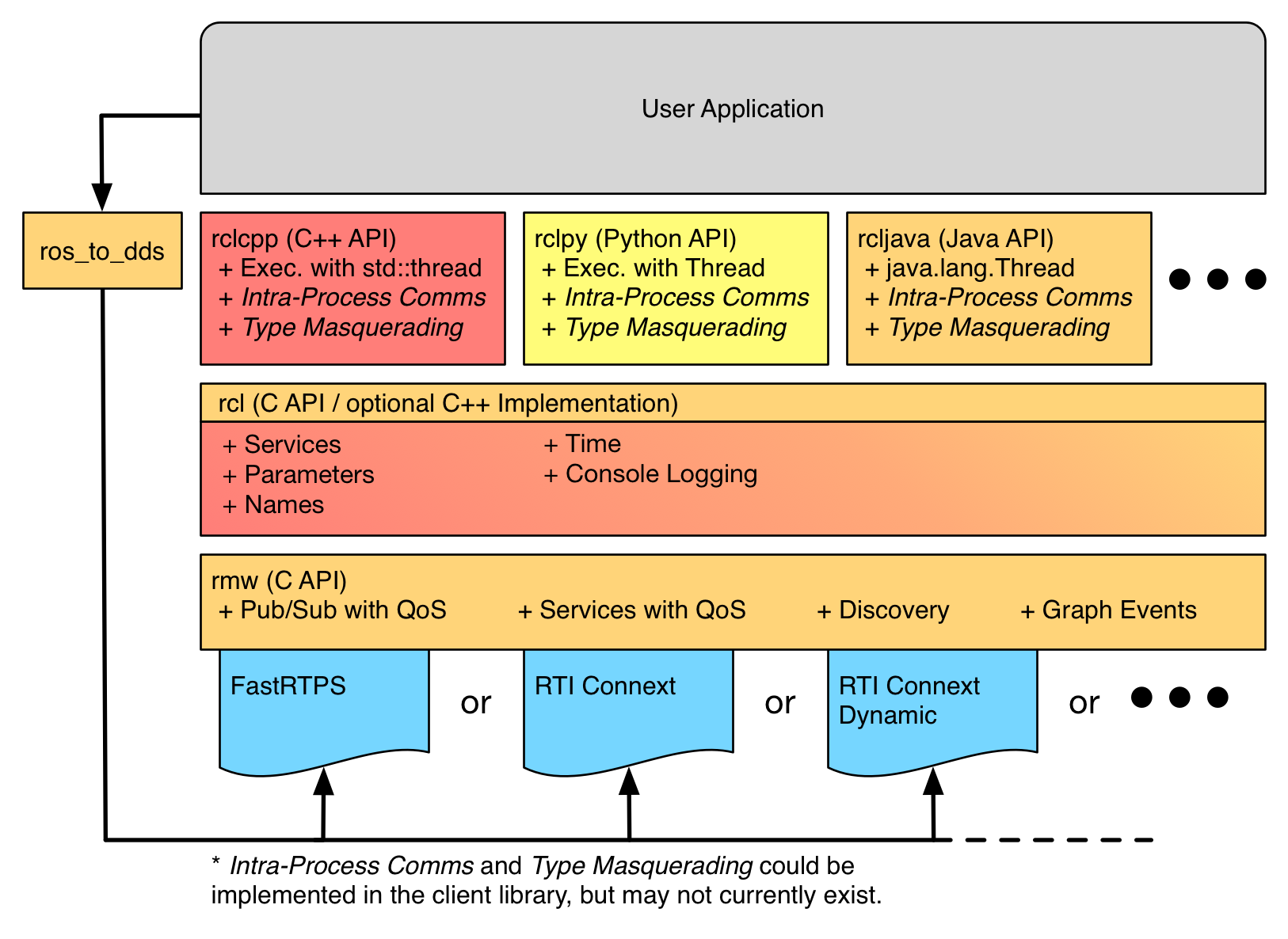}
    \caption{ROS2 Internal API Architecture~\cite{ROSArch}}
    \label{fig:rosArch}
\end{figure}

\begin{enumerate}
    \item The ROS middleware interface (RMW). This API provides an interface between ROS2 middleware and DDS or Real-Time Publish-Subscribe Interoperability Wire Protocol (RTPS), which are used to provide discovery, publish and subscribe services. This layer is implemented in C.
    \item The ROS client library interface (RCL). This layer is used to implement the client libraries, which provides access to the ROS graph based on concepts like Topics, Services, and Actions. RCL may come in a variety of programming languages such as Java, Python, and C++.
\end{enumerate}

According to Figure~\ref{fig:rosArch}, the middleware layer is developed using C, while the higher layer can be implemented using various programming languages. This diversity of programming languages hardens the analysis of the ROS2 framework and challenges existing techniques. Another challenge imposed by this architecture is the usage of other third party projects such as DDS. Therefore, conducting a holistic analysis is required to validate the reliability and security of ROS2. Static analysis methods provide an adequate approach to achieve this goal~\cite{Santos2016}.

However, software evaluation in the robotics domain needs special requirements and guidelines\cite{Santos2016}. For instance, in the context of safety-critical systems (e.g. automotive), enforcement of coding standards is a necessity~\cite{Andre20}. These standards consist of rules and guidelines that aim to increase the robustness and reliability of software systems. They impose style guidelines and restrict the usage of certain programming languages’ features to avoid common mistakes. Static analysis facilitates the detection of violations of these rules. Therefore, conducting static analysis for ROS2 is feasible, because the ROS community uses Git repositories for its source code, hosting them freely on GitHub. 

\section{Overall Approach}\label{sec:overallApproach}
This section explains our approach to achieve the two contributions of this work. We follow the workflow illustrated in Figure~\ref{fig:approach}. Following is a description of each step:

\begin{figure}[h!]
    \centering
    \includegraphics[width=1\columnwidth]{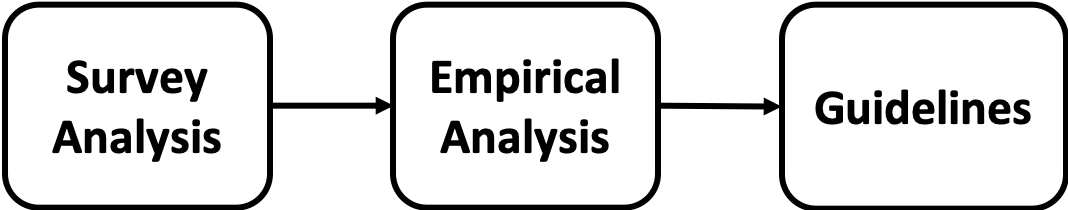}
    \caption{Analysis Workflow.}
    \label{fig:approach}
\end{figure}

\begin{enumerate}
  \item \textbf{Conducting survey analysis}. This step aims to: (a) understand the software architecture of the ROS2 framework and (b) identify the main focus and limitation of prior work that target analyzing ROS framework. Specifically, we perform a small-scale systematic literature review by considering papers recently published in software engineering conferences (i.e. ICSE and ASE), security conferences (i.e. CCS and USENIX), and robotics conferences (i.e. IROS and ICRA). The process of conducting the survey analysis and the results are discussed in Section~\ref{sec:survey}.
  
  \item \textbf{Empirical analysis}. In this step, we focus on analyzing ROS2 source code based on the identified directions in the survey analysis step. Specifically, We perform a static analysis using bug detection tools that are available in the literature such as PMD~\cite{pmd}. This step provides an overall picture of the software quality of the ROS2 code. The static analysis tool selection and the experimental approach are discussed in Section~\ref{sec:approachExp} and the results of the empirical study are presented in Section~\ref{sec:empirical}. 
  
  \item \textbf{Providing Guidelines}. In this step, we discuss interesting observations based on the conducted analysis in the survey analysis and empirical study. Finally, this step describes future research directions in this area (Section~\ref{sec:disc}).
\end{enumerate}

\section{Emperical Study}\label{sec:approachExp}
This section describes our emprical approach for assessing the software quality of ROS2. It introduces the selection criteria for selecting the static analysis tool for detecting bugs and identifies the software properties that we verified. This section also describes the collected software corpus that represents ROS2 Java projects.

\subsection{Static Analysis Tool Selection}
We define two selection criteria for identifying the optimal tool to develop our detection rules. The tool should be:
\begin{enumerate}
    \item open source and is still actively supported by the community.
    \item easy-to-use and facilitating the integration of new rules.
\end{enumerate}

To this end, we identified several open-source static analysis tools that are presented for detecting bugs in Java programs~\cite{Alhanahnah2018}. The tools are as follow: 

\begin{enumerate}
    \item FindBugs~\cite{findbugs}: is an open-source tool for detecting bugs in Java code. It is a static analysis tool on Java bytecode and can be used via command line and integrated into different IDEs. FindBugs can discover various types of bugs including problematic coding practices and vulnerabilities. FindBugs rules can be created using the Visitor pattern (Java API). 
    
    \item Hammurapi~\cite{hammurapi}: is an open-source tool for analyzing Java source code. It can be integrated into IDEs and is developed with scalability in mind. Hammurapi employs Abstract Syntax Tree (AST), where new rules can be added to this tool, using java code or XML rules. However, this tool is rather complicated~\cite{aderhold2013}, which is driven by the fact that Hammurapi consists to several components, including web services and database that should be downloading to use this tool.
    
    \item Jlint~\cite{Jlint}: is written in C++ for detecting common programming errors in Java (e.g., race condition). Jlint performs semantic and syntax analysis on Java bytecode for accomplishing its duties. Although new rules can be integrated into Jlint, it will require modifying Jlint's source code~\cite{aderhold2013}, which makes Jlint difficult to expand.
    
    \item PMD: is an open-source tool, which is written in Java and it checks Java source code for a set of predefined bugs. PMD can be used through the command line, and graphical user interface via the available plugins for various IDEs. PMD constructs the AST, and then examines the constructed AST for detecting bugs. PMD rules can be defined using Java code (Visitor pattern) or XPath queries. This provides more flexibility and makes it easier for an extension.
\end{enumerate}

\noindent\textbf{Other Tools.} Although the focus of our analysis is on Java ROS projects, we also explored tools for analyzing C++ code. One of the off-the-shelf tools that are used for analyzing C++ ROS code is Cppcheck~\cite{Cppcheck}. This tool has been used in prior work to perform security analysis~\cite{kim2018security} and collect code metrics~\cite{Santos2016}.

Accordingly, PMD has been selected for performing our experiment, because PMD is an open-source tool and can be easily expanded with new rule sets. Unlike other tools that require changing the source codes of the tools, or are limited to a specific method for adding new rules, PMD is flexible, easy-to-use, and deemed as a cross-architectural analysis tool, as it can analyze different programming languages. PMD also covers different categories of bug detection (i.e. security, code style and compliance with code quality)

\subsection{Empirical Study Design}
\begin{figure}[h!]
    \centering
    \includegraphics[width=0.9\columnwidth]{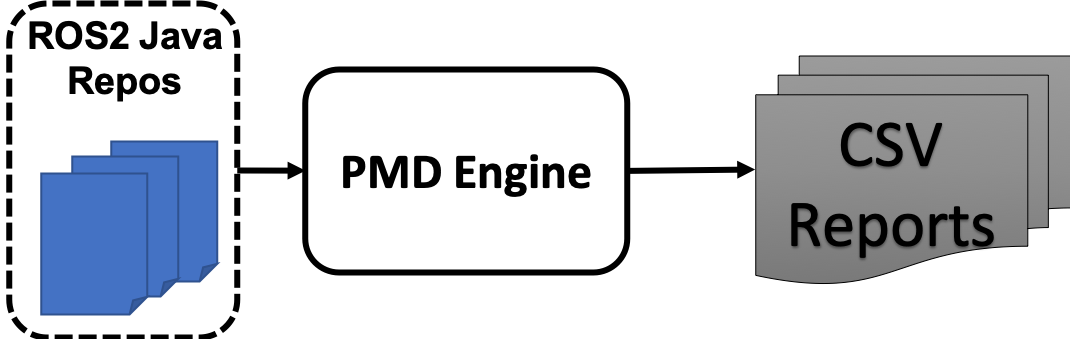}
    \caption{Static Analysis Using PMD}
    \label{fig:approachPMD}
\end{figure}

This section describes our approach for analyzing ROS software quality. In our experiment, we leverage PMD to check several properties of ROS2 Java projects, as illustrated in Figure~\ref{fig:approachPMD}. 

\noindent\textbf{Software Corpus:} to find Java ROS2 code, we search the official website of ROS2~\footnote{https://index.ros.org/doc/ros2/}, which provided us the Github links for accessing the ROS2 Java projects. ROS2 Java code includes Android projects. Overall, we found eight Github projects.

\noindent\textbf{PMD Rules:} as mentioned earlier, PMD has built-in rules for analyzing various code properties. In this experiment, we use the rules summarized in Table~\ref{tab:PMD-Rules}. PMD is inadequate to verify security properties of ROS2 because it only has two security rules, which are limited to: (1) check hardcoded values for cryptographic operations, and (2) identify hardcoded initialization vector in cryptographic operations. However, PMD rules can be extended to involve additional weaknesses and code smell ~\cite{Alhanahnah2018}.

\begin{table}[h!]
    \centering
    \caption{Summary of PMD Rules}
    \begin{tabular}{|p{6cm}|c|}
    \hline
        Category & \# of Rules \\\hline
        Best Practices & $50$ \\\hline
        Code Style & $62$ \\\hline
        Design & $46$ \\\hline
        Error Prone & $99$ \\\hline
        Performance & $30$ \\\hline
        Security & $2$ \\\hline
        Additional rulesets (Android \& Security Code Guidelines) & $5$ \\\hline
    \end{tabular}
    \label{tab:PMD-Rules}
\end{table}

\section{Results}\label{sec:study}
This section discusses the results of the two contributions of this work, the literature review and empirical study. 
\subsection{Survey Analysis} \label{sec:survey}
    We conducted a literature review to find publications related to ROS. Specifically, we are seeking related work that leverages program analysis techniques for assessing the software features of ROS. Table~\ref{tab:rw} presents the results of our systematic approach to exploring the literature. We research several research communities since program analysis efforts of robot software fall under several domains, including software engineering, security, and robotics. we consider publications in top venues in these three communities. The total number of publications is $16$, obtained mainly from robotics venues. Although our literature analysis is not comprehensive, it helps us to draw interesting conclusions. For instance, the absence of publications in communities such as software engineering and security is an interesting observation and requires more investigation. This observation shows great opportunities to improve the software features (i.e. safety, security, and reliability) of ROS.

    \begin{table}[h!]
        \centering
        \caption{Summary of Related Work}
        \begin{tabular}{|p{1.5cm}|p{1.8cm}|p{1.4cm}|c|}
        \hline
            Domain & Venue & Years & \# of papers \\\hline
            \multirow{4}{2.3cm}{Software Engineering }& ICSE  & \multirow{4}{*}{2017 - 2019} & $0$ \\\cline{2-2}\cline{4-4}
            & ESEC/FSE & & \cite{Kate2018} \\\cline{2-2}\cline{4-4}
            & ISSTA & & \cite{Ore2017FP} \\\cline{2-2}\cline{4-4}
            & ASE & & 0\\\hline
            \multirow{3}{*}{Security} & CCS & \multirow{3}{*}{2017 - 2019} & \cite{Choi2018}$^*$ \\\cline{2-2}\cline{4-4}
            & USENIX & & 0 \\\cline{2-2}\cline{4-4}
            & NDSS & & 0 \\\hline
            \multirow{3}{*}{Robotics} & IROS & \multirow{3}{*}{2016 - 2019} & \cite{Dieber2016,Santos2016,Ore2017IROS} \\\cline{2-2}\cline{4-4}
            & ICRA & & \cite{DeMarinis2019,quigley2009ros,White2019} \\\cline{2-2}\cline{4-4}
            & Other & & \cite{kim2018security,McClean2013,Breiling2017,cerrudo2017hacking,Dieber2020,Caiazza2019,Giaretta2018} \\\hline
        \end{tabular}
        \footnotesize{\\$^*$: in robotics domain but not ROS related}\\
        \label{tab:rw}
    \end{table}
    
    We further classified the identified publications in the literature based on the (1) utilized approach static or dynamic, and (2) the goal of the analysis. The goals are divided into security, quality, and physical units. We excluded the papers that do not provide or utilize a tool for assessing ROS like ~\cite{quigley2009ros,Caiazza2019}. Although some papers such as~\cite{Choi2018} do not evaluate the ROS framework, we decided to include them in the classifications because they utilize ROS in the conducted evaluation. Table~\ref{tab:classification} presents our classification of the related work. We can conclude from this table that dynamic analysis is a popular approach to perform a security assessment, but static analysis is a common approach to investigate a certain software property. Another observation based on the results in Table~\ref{tab:classification}, the majority of prior work targets ROS1. We also notice that the majority of the dynamic analysis approaches are concentrated around the network layer than the application layer, except ROSPenTo and Roschaos~\cite{Dieber2020}, which exploit the usage of certain ROS APIs. 
    
    \begin{table}[h!]
        \centering
        \caption{Classification of the Related Work}
        \begin{tabular}{|c|p{1.5cm}|p{2.2cm}|c|}
        \hline
        Reference & Approach & Goal & Scope\\\hline
        PhrikyUnits~\cite{Ore2017,Ore2017FP,Ore2017IROS} & \multirow{3}{*}{Static} & \multirow{2}{3cm}{Inconsistency in physical units} & ROS1\\\cline{1-1} \cline{4-4}
        Phys\cite{Kate2018} &  & & ROS1\\\cline{1-1} \cline{3-4}
        HAROS\cite{Santos2016} & & Code Metrics & ROS1\\\cline{1-1}\cline{2-4} 
        \cite{Choi2018} & \multirow{6}{*}{Dynamic} & \multirow{6}{*}{Security} & --\\\cline{1-1}\cline{4-4}
        \cite{DeMarinis2019} &  & & ROS1\\\cline{1-1}\cline{4-4}
        \cite{Dieber2016} &  &  & ROS1\\\cline{1-1}\cline{4-4}
        \cite{White2019} &  &  & ROS1\\\cline{1-1}\cline{4-4}
        \cite{Breiling2017} &  &  & ROS1\\\cline{1-1}\cline{4-4}
        \cite{cerrudo2017hacking} &  & & ROS1\\\cline{1-1}\cline{4-4}
        \cite{McClean2013} &  &  & ROS1\\\cline{1-1}\cline{4-4}
        ROSPenTo~\cite{Dieber2020} &  &  & ROS1\\\cline{1-1}\cline{4-4}
        Roschaos~\cite{Dieber2020} &  & & ROS1\\\hline 
        \cite{kim2018security} & Static and Dynamic & Performance and Security & ROS2\\\hline 
        \end{tabular}
        \label{tab:classification}
    \end{table}

\subsection{Empirical Analysis Results}\label{sec:empirical}
\begin{figure}[h!]
    \centering
    \includegraphics[width=1\columnwidth]{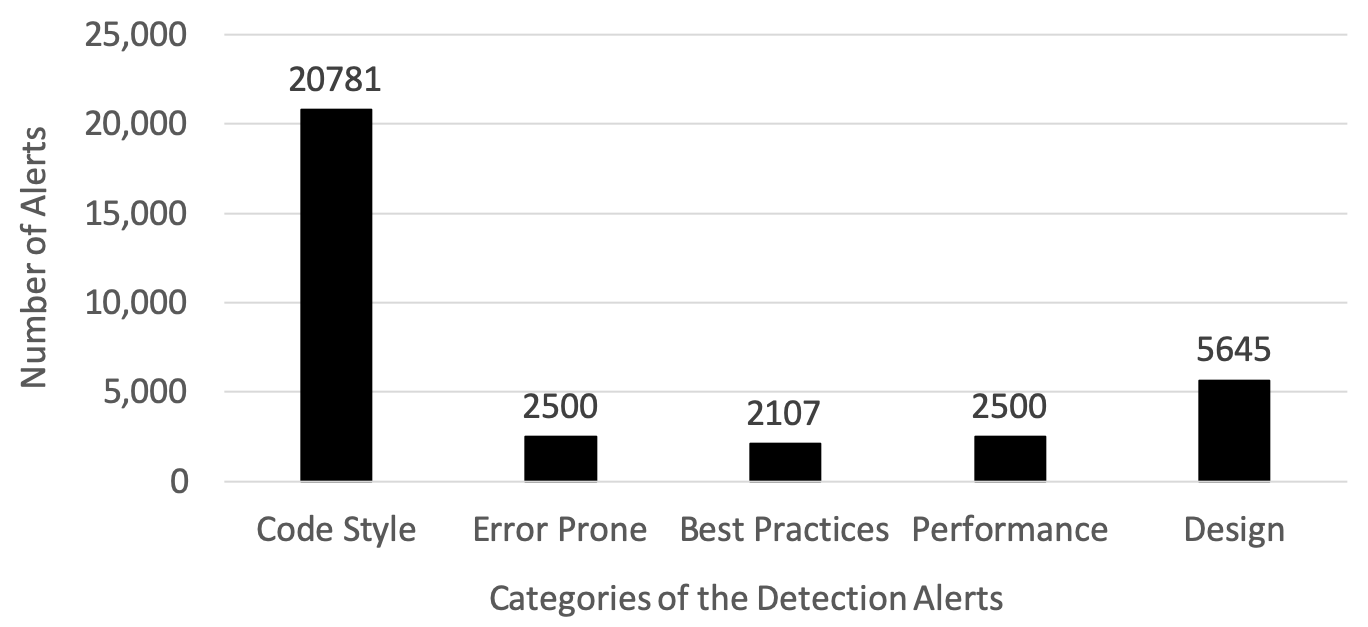}
    \caption{Distribution of all detected violations}
    \label{fig:DistributionOfErrors}
\end{figure}

Figure~\ref{fig:DistributionOfErrors} presents the results of the performed static analysis. The total number of alerts is 33,533 that have been detected by PMD, where $62\%$ of the alerts are \textit{code style} issues, while the lowest number of alerts ($6\%$) are under the category \textit{Best Practices}. These results show that the majority of the alerts are not dangerous. However, the \textit{Error Prone} and \textit{Performance} require further investigation. Our analysis reveals the existence of several empty control flow statements as summarized in Table~\ref{tab:empty}. This kind of code implementation can make the code look suspicious, as attackers employ empty blocks to bypass detection mechanisms~\cite{Wang2016}. 

\begin{table}[h!]
    \centering
    \caption{Summary of empty block}
    \begin{tabular}{|c|c|}
    \hline
        Rule & Count \\\hline
        Avoid empty while statements & 4 \\\hline
        Avoid empty if statements & 4 \\\hline
        Avoid empty catch blocks & 8 \\\hline
    \end{tabular}
    \label{tab:empty}
\end{table}

\begin{figure*}[h!]
    \centering
    \includegraphics[width=1.8\columnwidth]{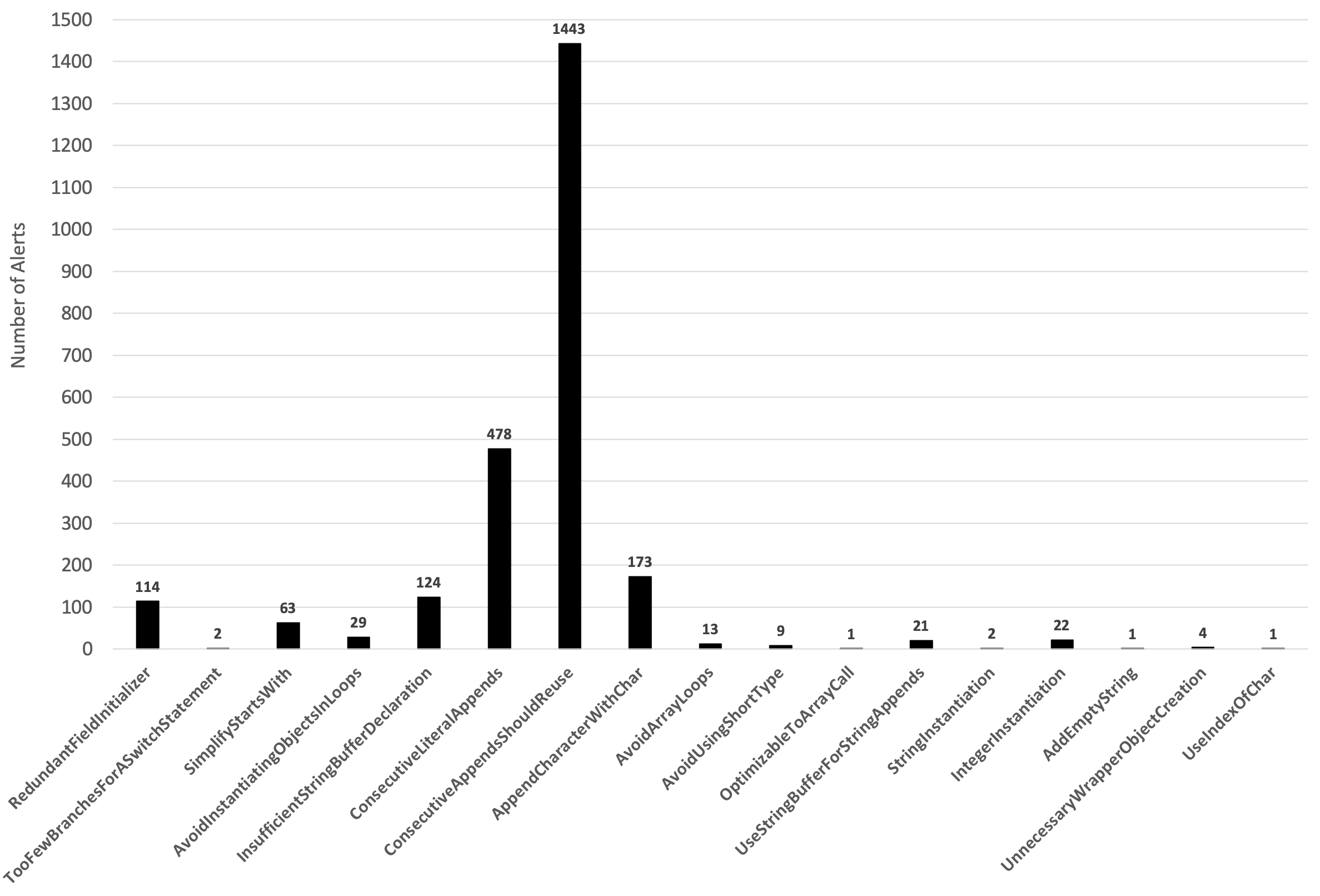}
    \caption{Distribution of Performance violations}
    \label{fig:DistributionOfPerformance}
\end{figure*}

Since performance can play a vital role in the robotics domain~\cite{kim2018security}, then performance issues should not exist in the ROS code. Although performance alerts constitute only $7\%$ of the overall violations, Figure~\ref{fig:DistributionOfPerformance} shows $17$ performance rules are violated. The goal of the performance alerts in PMD is to avoid coding practices that can affect the performance. For example, Java provide APIs such as \textit{asList} to enhance the performance; therefore, when the developer creates a list from an array of objects using a loop instead of \textit{asList}, then PMD will fire an alert accordingly.

We also manually investigate some of the results, with a particular focus on exploring security features in ROS2. Listing~\ref{lst:threat-assertions} shows some security flags that are set to false. For instance, the variable \textit{\_secureUseSecure} enables the secure communication between ROS nodes. This is a serious issue since this code snippet is located in Data Distribution Service (DDS), which is the extension that provides security capabilities to ROS2.
\begin{lstlisting}[caption=Code snippet from RTI Perftest~\protect\footnote{https://github.com/rticommunity/rtiperftest},label={lst:threat-assertions}]
    private boolean _secureUseSecure = false;
    private boolean _secureIsSigned = false;
    private boolean _secureIsDataEncrypted = false; // User Data
    private boolean _secureIsSMEncrypted = false; // Sub-message
    private boolean _secureIsDiscoveryEncrypted = false;
    private String _secureCertAuthorityFile = null;
    private String _secureCertificateFile = null;
    private String _securePrivateKeyFile = null;
\end{lstlisting}

\section{Related Work} \label{sec:rw}

\noindent\textbf{Survey Studies.} Prior work highlights several aspects of ROS~\cite{Caiazza2019,quigley2009ros}. Caiazza et al.~\cite{Caiazza2019} discuss approaches for enhancing the security of ROS, while Quigley et al. ~\cite{quigley2009ros} give an overview of ROS. However, the body of knowledge lack of a dedicated and systematic study that discusses tools and techniques proposed for evaluating ROS software. In this work, we attempt to cover this research gap by conducting a small-scale literature review. 

\noindent\textbf{ROS Code Analysis.}  ROS implementation has been evaluated from different angles. The security aspect was not part of ROS's goals, even security in ROS2 is overlooked, except providing a few optional security extensions~\cite{kim2018security}; therefore, ROS suffers from significant security weaknesses including plaintext communications, open ports, and unencrypted storage~\cite{McClean2013,Breiling2017,cerrudo2017hacking}. The state-of-the-art~\cite{Dieber2020,McClean2013,DeMarinis2019,White2019,Giaretta2018} show the communication between ROS nodes is a major concern. Therefore, many of the proposed solutions consider and demonstrated attacks target this limitation. For instance, several attacks have been demonstrated in the literature. Open source penetration testing tools have been leveraged to perform the attacks on a cyber-physical security honeypot developed on top of ROS~\cite{McClean2013}. The conducted attacks show that ROS framework suffers from major weaknesses, which allowed performing man-in-the-middle attacks~\cite{Callegati2009}. To overcome this challenge, a security layer is proposed to be deployed on top of the ROS framework in~\cite{Dieber2016}. This security layer provides integrity and confidentiality by maintaining an authorization server to enable secure communication between ROS nodes. Similarly, Breiling et al.,~\cite{Breiling2017} propose a secure communication channel for ROS to handle the communication between two nodes. Another experiment discovers over 100 publicly-accessible hosts running a ROS master, after scanning the whole IPv4 scope~\cite{DeMarinis2019}. 
Unlike these works, the scope of our analysis is considering ROS2, and we apply a static analysis approach for verifying the security properties of ROS2.

Another stream of research applies static analysis methods to (1) identify inconsistency in physical units~\cite{Kate2018,Ore2017IROS,Ore2017FP}, verify compliance with standard coding style and collect code matrices~\cite{Santos2016}, and (3) check security aspects~\cite{kim2018security}. In this work, we consider several software properties instead of focusing on certain properties. HAROS~\cite{Santos2016} is the closest work to our empirical study. HAROS uses a set of off-the-shelf tools and focuses on identifying code matrices in ROS1 code. In our work, we focus on analyzing ROS2 and considered a different set of software properties. For example, PMD supports collecting information about the code comments similar to HAROS, but we have not enabled this rule in our analysis, because we are interested in capturing more critical aspects. Finally, opposite to the aforementioned static techniques, our experiment targets ROS2 Java code.

\section{Discussion}\label{sec:disc}
We applied a systematic approach for collecting and analyzing the literature. We also show the research trends in the ROS domain. Static analysis approaches for assessing specific properties of ROS, while dynamic analysis approaches are used mainly for evaluating security aspects of ROS. However, our survey study is limited to 16 papers. Therefore, additional publications should be considered to provide useful insights. 

In the empirical study, we leveraged the default rules that come with PMD. However, these rules are not mandatory tailored for verify the requirements of ROS. Therefore, as for future work, PMD can be extended to detect other software issues~\cite{Alhanahnah2018}, which extends PMD to detect insecure implementation of the Secure Socket Layer (SSL) in Android applications. Specifically, more security rules need to be integrated into PMD, because the default security rule set in PMD covers only two issues. Furthermore, other issues such as detecting copy and paste can be also identified through PMD, which provides this feature for detecting copy and paste in several programming languages like C++ and Python. Therefore, the detection of copy and paste can help in identifying vulnerable patterns in ROS, thus preventing the propagation of insecure code.

\section{Conclusion}\label{sec:conclusion}
In this paper, we perform a small-scale systematic analysis to identify the work that addresses software quality concerns about ROS2 software. We show there is a tendency to apply a dynamic analysis approach for identifying security issues, while static analysis approaches are employed to check specific program properties and verify compliance with code style standards. This small-scale analysis shows a major shortcoming of large-scale systematic analysis work in this area. We also perform an experimental analysis to evaluate the software quality of ROS2. We leverage PMD, an open-source static analysis tool, to achieve this goal. Our analysis shows that the majority of the detected alerts belong to code style issues. PMD also detects error-prone and performance issues.

\bibliographystyle{ACM-Reference-Format}
\bibliography{ref}


\begin{thebibliography}{30}


\ifx \showCODEN    \undefined \def \showCODEN     #1{\unskip}     \fi
\ifx \showDOI      \undefined \def \showDOI       #1{#1}\fi
\ifx \showISBNx    \undefined \def \showISBNx     #1{\unskip}     \fi
\ifx \showISBNxiii \undefined \def \showISBNxiii  #1{\unskip}     \fi
\ifx \showISSN     \undefined \def \showISSN      #1{\unskip}     \fi
\ifx \showLCCN     \undefined \def \showLCCN      #1{\unskip}     \fi
\ifx \shownote     \undefined \def \shownote      #1{#1}          \fi
\ifx \showarticletitle \undefined \def \showarticletitle #1{#1}   \fi
\ifx \showURL      \undefined \def \showURL       {\relax}        \fi
\providecommand\bibfield[2]{#2}
\providecommand\bibinfo[2]{#2}
\providecommand\natexlab[1]{#1}
\providecommand\showeprint[2][]{arXiv:#2}

\bibitem[\protect\citeauthoryear{??}{ROS}{[n. d.]}]%
        {ROSArch}
 \bibinfo{year}{[n. d.]}\natexlab{}.
\newblock \bibinfo{title}{{Core Stack Developer Overview}}.
\newblock
  \bibinfo{howpublished}{\url{http://docs.ros2.org/dashing/developer_overview.html}}.
    (\bibinfo{year}{[n. d.]}).
\newblock
\newblock
\shownote{Accessed at Oct 03, 2019.}


\bibitem[\protect\citeauthoryear{??}{pmd}{[n. d.]}]%
        {pmd}
 \bibinfo{year}{[n. d.]}\natexlab{}.
\newblock \bibinfo{title}{{PMD Tool}}.
\newblock
  \bibinfo{howpublished}{\url{https://pmd.github.io/pmd-5.8.1/index.html}}.
  (\bibinfo{year}{[n. d.]}).
\newblock
\newblock
\shownote{Accessed at Oct 03, 2019.}


\bibitem[\protect\citeauthoryear{??}{fin}{2019}]%
        {findbugs}
 \bibinfo{year}{accessed at Nov. 2019}\natexlab{}.
\newblock \bibinfo{title}{FindBugs - Find Bugs in Java Programs}.
\newblock \bibinfo{howpublished}{\url{http://findbugs.sourceforge.net/}}.
  (\bibinfo{year}{accessed at Nov. 2019}).
\newblock


\bibitem[\protect\citeauthoryear{??}{ham}{2019}]%
        {hammurapi}
 \bibinfo{year}{accessed at Nov. 2019}\natexlab{}.
\newblock \bibinfo{title}{Hammurapi Group - Java tools and libraries}.
\newblock
  \bibinfo{howpublished}{\url{http://www.hammurapi.biz/hammurapi-biz/ef/xmenu/hammurapi-group/products/hammurapi/index.html}}.
    (\bibinfo{year}{accessed at Nov. 2019}).
\newblock


\bibitem[\protect\citeauthoryear{??}{Jli}{2019}]%
        {Jlint}
 \bibinfo{year}{accessed at Nov. 2019}\natexlab{}.
\newblock \bibinfo{title}{Jlint - Find Bugs in Java Programs}.
\newblock \bibinfo{howpublished}{\url{http://jlint.sourceforge.net}}.
  (\bibinfo{year}{accessed at Nov. 2019}).
\newblock


\bibitem[\protect\citeauthoryear{Aderhold and Kochtchi}{Aderhold and
  Kochtchi}{2013}]%
        {aderhold2013}
\bibfield{author}{\bibinfo{person}{Markus Aderhold} {and}
  \bibinfo{person}{Artjom Kochtchi}.} \bibinfo{year}{2013}\natexlab{}.
\newblock \bibinfo{title}{Tailoring PMD to Secure Coding}.
\newblock \bibinfo{howpublished}{Tech. Rep.}.   (\bibinfo{year}{2013}).
\newblock


\bibitem[\protect\citeauthoryear{{Alhanahnah} and {Yan}}{{Alhanahnah} and
  {Yan}}{2018}]%
        {Alhanahnah2018}
\bibfield{author}{\bibinfo{person}{M. {Alhanahnah}} {and} \bibinfo{person}{Q.
  {Yan}}.} \bibinfo{year}{2018}\natexlab{}.
\newblock \showarticletitle{Towards best secure coding practice for
  implementing SSL/TLS}. In \bibinfo{booktitle}{{\em IEEE INFOCOM 2018 - IEEE
  Conference on Computer Communications Workshops (INFOCOM WKSHPS)}}.
  \bibinfo{pages}{1--6}.
\newblock
\showISSN{null}
\showDOI{%
\url{https://doi.org/10.1109/INFCOMW.2018.8407011}}


\bibitem[\protect\citeauthoryear{{Breiling}, {Dieber}, and
  {Schartner}}{{Breiling} et~al\mbox{.}}{2017}]%
        {Breiling2017}
\bibfield{author}{\bibinfo{person}{B. {Breiling}}, \bibinfo{person}{B.
  {Dieber}}, {and} \bibinfo{person}{P. {Schartner}}.}
  \bibinfo{year}{2017}\natexlab{}.
\newblock \showarticletitle{Secure communication for the robot operating
  system}. In \bibinfo{booktitle}{{\em 2017 Annual IEEE International Systems
  Conference (SysCon)}}. \bibinfo{pages}{1--6}.
\newblock


\bibitem[\protect\citeauthoryear{Caiazza, White, and Cortesi}{Caiazza
  et~al\mbox{.}}{2019}]%
        {Caiazza2019}
\bibfield{author}{\bibinfo{person}{Gianluca Caiazza}, \bibinfo{person}{Ruffin
  White}, {and} \bibinfo{person}{Agostino Cortesi}.}
  \bibinfo{year}{2019}\natexlab{}.
\newblock \bibinfo{booktitle}{{\em Enhancing Security in ROS}}.
\newblock \bibinfo{publisher}{Springer Singapore},
  \bibinfo{address}{Singapore}, \bibinfo{pages}{3--15}.
\newblock


\bibitem[\protect\citeauthoryear{{Callegati}, {Cerroni}, and
  {Ramilli}}{{Callegati} et~al\mbox{.}}{2009}]%
        {Callegati2009}
\bibfield{author}{\bibinfo{person}{F. {Callegati}}, \bibinfo{person}{W.
  {Cerroni}}, {and} \bibinfo{person}{M. {Ramilli}}.}
  \bibinfo{year}{2009}\natexlab{}.
\newblock \showarticletitle{Man-in-the-Middle Attack to the HTTPS Protocol}.
\newblock \bibinfo{journal}{{\em IEEE Security Privacy\/}} \bibinfo{volume}{7},
  \bibinfo{number}{1} (\bibinfo{date}{Jan} \bibinfo{year}{2009}),
  \bibinfo{pages}{78--81}.
\newblock


\bibitem[\protect\citeauthoryear{Cerrudo and Apa}{Cerrudo and Apa}{2017}]%
        {cerrudo2017hacking}
\bibfield{author}{\bibinfo{person}{Cesar Cerrudo} {and} \bibinfo{person}{Lucas
  Apa}.} \bibinfo{year}{2017}\natexlab{}.
\newblock \showarticletitle{Hacking robots before skynet}.
\newblock \bibinfo{journal}{{\em IOActive Website\/}} (\bibinfo{year}{2017}).
\newblock


\bibitem[\protect\citeauthoryear{Choi, Lee, Aafer, Fei, Tu, Zhang, Xu, and
  Deng}{Choi et~al\mbox{.}}{2018}]%
        {Choi2018}
\bibfield{author}{\bibinfo{person}{Hongjun Choi}, \bibinfo{person}{Wen-Chuan
  Lee}, \bibinfo{person}{Yousra Aafer}, \bibinfo{person}{Fan Fei},
  \bibinfo{person}{Zhan Tu}, \bibinfo{person}{Xiangyu Zhang},
  \bibinfo{person}{Dongyan Xu}, {and} \bibinfo{person}{Xinyan Deng}.}
  \bibinfo{year}{2018}\natexlab{}.
\newblock \showarticletitle{Detecting Attacks Against Robotic Vehicles: A
  Control Invariant Approach}. In \bibinfo{booktitle}{{\em Proceedings of the
  2018 ACM SIGSAC Conference on Computer and Communications Security}} {\em
  (\bibinfo{series}{CCS '18})}. \bibinfo{publisher}{ACM}, \bibinfo{address}{New
  York, NY, USA}, \bibinfo{pages}{801--816}.
\newblock
\showISBNx{978-1-4503-5693-0}
\showDOI{%
\url{https://doi.org/10.1145/3243734.3243752}}


\bibitem[\protect\citeauthoryear{{DeMarinis}, {Tellex}, {Kemerlis},
  {Konidaris}, and {Fonseca}}{{DeMarinis} et~al\mbox{.}}{2019}]%
        {DeMarinis2019}
\bibfield{author}{\bibinfo{person}{N. {DeMarinis}}, \bibinfo{person}{S.
  {Tellex}}, \bibinfo{person}{V.~P. {Kemerlis}}, \bibinfo{person}{G.
  {Konidaris}}, {and} \bibinfo{person}{R. {Fonseca}}.}
  \bibinfo{year}{2019}\natexlab{}.
\newblock \showarticletitle{Scanning the Internet for ROS: A View of Security
  in Robotics Research}. In \bibinfo{booktitle}{{\em 2019 International
  Conference on Robotics and Automation (ICRA)}}. \bibinfo{pages}{8514--8521}.
\newblock


\bibitem[\protect\citeauthoryear{{Dieber}, {Kacianka}, {Rass}, and
  {Schartner}}{{Dieber} et~al\mbox{.}}{2016}]%
        {Dieber2016}
\bibfield{author}{\bibinfo{person}{B. {Dieber}}, \bibinfo{person}{S.
  {Kacianka}}, \bibinfo{person}{S. {Rass}}, {and} \bibinfo{person}{P.
  {Schartner}}.} \bibinfo{year}{2016}\natexlab{}.
\newblock \showarticletitle{Application-level security for ROS-based
  applications}. In \bibinfo{booktitle}{{\em 2016 IEEE/RSJ International
  Conference on Intelligent Robots and Systems (IROS)}}.
  \bibinfo{pages}{4477--4482}.
\newblock


\bibitem[\protect\citeauthoryear{Dieber, White, Taurer, Breiling, Caiazza,
  Christensen, and Cortesi}{Dieber et~al\mbox{.}}{2020}]%
        {Dieber2020}
\bibfield{author}{\bibinfo{person}{Bernhard Dieber}, \bibinfo{person}{Ruffin
  White}, \bibinfo{person}{Sebastian Taurer}, \bibinfo{person}{Benjamin
  Breiling}, \bibinfo{person}{Gianluca Caiazza}, \bibinfo{person}{Henrik
  Christensen}, {and} \bibinfo{person}{Agostino Cortesi}.}
  \bibinfo{year}{2020}\natexlab{}.
\newblock \bibinfo{booktitle}{{\em Penetration Testing ROS}}.
\newblock \bibinfo{publisher}{Springer International Publishing},
  \bibinfo{address}{Cham}, \bibinfo{pages}{183--225}.
\newblock


\bibitem[\protect\citeauthoryear{Erős, Dahl, Bengtsson, Hanna, and
  Falkman}{Erős et~al\mbox{.}}{2019}]%
        {ers2019ros2}
\bibfield{author}{\bibinfo{person}{Endre Erős}, \bibinfo{person}{Martin Dahl},
  \bibinfo{person}{Kristofer Bengtsson}, \bibinfo{person}{Atieh Hanna}, {and}
  \bibinfo{person}{Petter Falkman}.} \bibinfo{year}{2019}\natexlab{}.
\newblock \bibinfo{title}{A ROS2 based communication architecture for control
  in collaborative and intelligent automation systems}.
\newblock   (\bibinfo{year}{2019}).
\newblock
\showeprint[arxiv]{cs.RO/1905.09654}


\bibitem[\protect\citeauthoryear{Giaretta, De~Donno, and Dragoni}{Giaretta
  et~al\mbox{.}}{2018}]%
        {Giaretta2018}
\bibfield{author}{\bibinfo{person}{Alberto Giaretta}, \bibinfo{person}{Michele
  De~Donno}, {and} \bibinfo{person}{Nicola Dragoni}.}
  \bibinfo{year}{2018}\natexlab{}.
\newblock \showarticletitle{Adding Salt to Pepper: A Structured Security
  Assessment over a Humanoid Robot}. In \bibinfo{booktitle}{{\em Proceedings of
  the 13th International Conference on Availability, Reliability and Security}}
  {\em (\bibinfo{series}{ARES 2018})}. \bibinfo{pages}{22:1--22:8}.
\newblock
\showISBNx{978-1-4503-6448-5}


\bibitem[\protect\citeauthoryear{Goforth}{Goforth}{[n. d.]}]%
        {Andre20}
\bibfield{author}{\bibinfo{person}{Andre Goforth}.} \bibinfo{year}{[n.
  d.]}\natexlab{}.
\newblock \bibinfo{booktitle}{{\em The Role and Impact of Software Coding
  Standards On System Integrity}}.
\newblock
\showDOI{%
\url{https://doi.org/10.2514/6.2013-5222}}
\showeprint{https://arc.aiaa.org/doi/pdf/10.2514/6.2013-5222}


\bibitem[\protect\citeauthoryear{Ingibergsson, Schultz, and
  Kuhrmann}{Ingibergsson et~al\mbox{.}}{2015}]%
        {Ingibergsson2015}
\bibfield{author}{\bibinfo{person}{Johann Thor~Mogensen Ingibergsson},
  \bibinfo{person}{Ulrik~Pagh Schultz}, {and} \bibinfo{person}{Marco
  Kuhrmann}.} \bibinfo{year}{2015}\natexlab{}.
\newblock \showarticletitle{On the Use of Safety Certification Practices in
  Autonomous Field Robot Software Development: A Systematic Mapping Study}. In
  \bibinfo{booktitle}{{\em Product-Focused Software Process Improvement}},
  \bibfield{editor}{\bibinfo{person}{Pekka Abrahamsson}, \bibinfo{person}{Luis
  Corral}, \bibinfo{person}{Markku Oivo}, {and} \bibinfo{person}{Barbara
  Russo}} (Eds.). \bibinfo{publisher}{Springer International Publishing},
  \bibinfo{address}{Cham}, \bibinfo{pages}{335--352}.
\newblock
\showISBNx{978-3-319-26844-6}


\bibitem[\protect\citeauthoryear{Kate, Ore, Zhang, Elbaum, and Xu}{Kate
  et~al\mbox{.}}{2018}]%
        {Kate2018}
\bibfield{author}{\bibinfo{person}{Sayali Kate}, \bibinfo{person}{John-Paul
  Ore}, \bibinfo{person}{Xiangyu Zhang}, \bibinfo{person}{Sebastian Elbaum},
  {and} \bibinfo{person}{Zhaogui Xu}.} \bibinfo{year}{2018}\natexlab{}.
\newblock \showarticletitle{Phys: Probabilistic Physical Unit Assignment and
  Inconsistency Detection}. In \bibinfo{booktitle}{{\em Proceedings of the 2018
  26th ACM Joint Meeting on European Software Engineering Conference and
  Symposium on the Foundations of Software Engineering}} {\em
  (\bibinfo{series}{ESEC/FSE 2018})}. \bibinfo{publisher}{ACM},
  \bibinfo{address}{New York, NY, USA}, \bibinfo{pages}{563--573}.
\newblock
\showISBNx{978-1-4503-5573-5}
\showDOI{%
\url{https://doi.org/10.1145/3236024.3236035}}


\bibitem[\protect\citeauthoryear{Kim, Smereka, Cheung, Nepal, and Grobler}{Kim
  et~al\mbox{.}}{2018}]%
        {kim2018security}
\bibfield{author}{\bibinfo{person}{Jongkil Kim}, \bibinfo{person}{Jonathon~M
  Smereka}, \bibinfo{person}{Calvin Cheung}, \bibinfo{person}{Surya Nepal},
  {and} \bibinfo{person}{Marthie Grobler}.} \bibinfo{year}{2018}\natexlab{}.
\newblock \showarticletitle{Security and performance considerations in ros 2: A
  balancing act}.
\newblock \bibinfo{journal}{{\em arXiv preprint arXiv:1809.09566\/}}
  (\bibinfo{year}{2018}).
\newblock


\bibitem[\protect\citeauthoryear{Marjamaki}{Marjamaki}{2019}]%
        {Cppcheck}
\bibfield{author}{\bibinfo{person}{D. Marjamaki}.} \bibinfo{year}{accessed at
  Nov. 2019}\natexlab{}.
\newblock \bibinfo{title}{Cppcheck - A tool for static C/C++ code analysis}.
\newblock \bibinfo{howpublished}{\url{http://cppcheck.sourceforge.net/}}.
  (\bibinfo{year}{accessed at Nov. 2019}).
\newblock


\bibitem[\protect\citeauthoryear{McClean, Stull, Farrar, and
  Mascareñas}{McClean et~al\mbox{.}}{2013}]%
        {McClean2013}
\bibfield{author}{\bibinfo{person}{Jarrod McClean},
  \bibinfo{person}{Christopher Stull}, \bibinfo{person}{Charles Farrar}, {and}
  \bibinfo{person}{David Mascareñas}.} \bibinfo{year}{2013}\natexlab{}.
\newblock \showarticletitle{{A preliminary cyber-physical security assessment
  of the Robot Operating System (ROS)}}. In \bibinfo{booktitle}{{\em Unmanned
  Systems Technology XV}}, \bibfield{editor}{\bibinfo{person}{Robert~E.
  Karlsen}, \bibinfo{person}{Douglas~W. Gage}, \bibinfo{person}{Charles~M.
  Shoemaker}, {and} \bibinfo{person}{Grant~R. Gerhart}} (Eds.),
  Vol.~\bibinfo{volume}{8741}. International Society for Optics and Photonics,
  \bibinfo{publisher}{SPIE}, \bibinfo{pages}{341 -- 348}.
\newblock


\bibitem[\protect\citeauthoryear{{Ore}, {Elbaum}, and {Detweiler}}{{Ore}
  et~al\mbox{.}}{2017}]%
        {Ore2017IROS}
\bibfield{author}{\bibinfo{person}{J. {Ore}}, \bibinfo{person}{S. {Elbaum}},
  {and} \bibinfo{person}{C. {Detweiler}}.} \bibinfo{year}{2017}\natexlab{}.
\newblock \showarticletitle{Dimensional inconsistencies in code and ROS
  messages: A study of 5.9M lines of code}. In \bibinfo{booktitle}{{\em 2017
  IEEE/RSJ International Conference on Intelligent Robots and Systems (IROS)}}.
  \bibinfo{pages}{712--718}.
\newblock
\showISSN{2153-0866}
\showDOI{%
\url{https://doi.org/10.1109/IROS.2017.8202229}}


\bibitem[\protect\citeauthoryear{Ore, Detweiler, and Elbaum}{Ore
  et~al\mbox{.}}{2017a}]%
        {Ore2017FP}
\bibfield{author}{\bibinfo{person}{John-Paul Ore}, \bibinfo{person}{Carrick
  Detweiler}, {and} \bibinfo{person}{Sebastian Elbaum}.}
  \bibinfo{year}{2017}\natexlab{a}.
\newblock \showarticletitle{Lightweight Detection of Physical Unit
  Inconsistencies Without Program Annotations}. In \bibinfo{booktitle}{{\em
  Proceedings of the 26th ACM SIGSOFT International Symposium on Software
  Testing and Analysis}} {\em (\bibinfo{series}{ISSTA 2017})}.
  \bibinfo{publisher}{ACM}, \bibinfo{address}{New York, NY, USA},
  \bibinfo{pages}{341--351}.
\newblock
\showISBNx{978-1-4503-5076-1}
\showDOI{%
\url{https://doi.org/10.1145/3092703.3092722}}


\bibitem[\protect\citeauthoryear{Ore, Detweiler, and Elbaum}{Ore
  et~al\mbox{.}}{2017b}]%
        {Ore2017}
\bibfield{author}{\bibinfo{person}{John-Paul Ore}, \bibinfo{person}{Carrick
  Detweiler}, {and} \bibinfo{person}{Sebastian Elbaum}.}
  \bibinfo{year}{2017}\natexlab{b}.
\newblock \showarticletitle{Phriky-units: A Lightweight, Annotation-free
  Physical Unit Inconsistency Detection Tool}. In \bibinfo{booktitle}{{\em
  Proceedings of the 26th ACM SIGSOFT International Symposium on Software
  Testing and Analysis}} {\em (\bibinfo{series}{ISSTA 2017})}.
  \bibinfo{publisher}{ACM}, \bibinfo{address}{New York, NY, USA},
  \bibinfo{pages}{352--355}.
\newblock
\showISBNx{978-1-4503-5076-1}
\showDOI{%
\url{https://doi.org/10.1145/3092703.3098219}}


\bibitem[\protect\citeauthoryear{Quigley, Conley, Gerkey, Faust, Foote, Leibs,
  Wheeler, and Ng}{Quigley et~al\mbox{.}}{2009}]%
        {quigley2009ros}
\bibfield{author}{\bibinfo{person}{Morgan Quigley}, \bibinfo{person}{Ken
  Conley}, \bibinfo{person}{Brian Gerkey}, \bibinfo{person}{Josh Faust},
  \bibinfo{person}{Tully Foote}, \bibinfo{person}{Jeremy Leibs},
  \bibinfo{person}{Rob Wheeler}, {and} \bibinfo{person}{Andrew~Y Ng}.}
  \bibinfo{year}{2009}\natexlab{}.
\newblock \showarticletitle{ROS: an open-source Robot Operating System}. In
  \bibinfo{booktitle}{{\em ICRA workshop on open source software}},
  Vol.~\bibinfo{volume}{3}. Kobe, Japan, \bibinfo{pages}{5}.
\newblock


\bibitem[\protect\citeauthoryear{{Santos}, {Cunha}, {Macedo}, and
  {Lourenço}}{{Santos} et~al\mbox{.}}{2016}]%
        {Santos2016}
\bibfield{author}{\bibinfo{person}{A. {Santos}}, \bibinfo{person}{A. {Cunha}},
  \bibinfo{person}{N. {Macedo}}, {and} \bibinfo{person}{C. {Lourenço}}.}
  \bibinfo{year}{2016}\natexlab{}.
\newblock \showarticletitle{A framework for quality assessment of ROS
  repositories}. In \bibinfo{booktitle}{{\em 2016 IEEE/RSJ International
  Conference on Intelligent Robots and Systems (IROS)}}.
  \bibinfo{pages}{4491--4496}.
\newblock
\showDOI{%
\url{https://doi.org/10.1109/IROS.2016.7759661}}


\bibitem[\protect\citeauthoryear{{Wang}, {Wang}, {Ming}, {Jiang}, and
  {Wu}}{{Wang} et~al\mbox{.}}{2016}]%
        {Wang2016}
\bibfield{author}{\bibinfo{person}{P. {Wang}}, \bibinfo{person}{S. {Wang}},
  \bibinfo{person}{J. {Ming}}, \bibinfo{person}{Y. {Jiang}}, {and}
  \bibinfo{person}{D. {Wu}}.} \bibinfo{year}{2016}\natexlab{}.
\newblock \showarticletitle{Translingual Obfuscation}. In
  \bibinfo{booktitle}{{\em 2016 IEEE European Symposium on Security and Privacy
  (EuroS P)}}. \bibinfo{pages}{128--144}.
\newblock
\showISSN{null}
\showDOI{%
\url{https://doi.org/10.1109/EuroSP.2016.21}}


\bibitem[\protect\citeauthoryear{White, Caiazza, Christensen, and
  Cortesi}{White et~al\mbox{.}}{2019}]%
        {White2019}
\bibfield{author}{\bibinfo{person}{Ruffin White}, \bibinfo{person}{Gianluca
  Caiazza}, \bibinfo{person}{Henrik Christensen}, {and}
  \bibinfo{person}{Agostino Cortesi}.} \bibinfo{year}{2019}\natexlab{}.
\newblock \bibinfo{booktitle}{{\em SROS1: Using and Developing Secure ROS1
  Systems}}.
\newblock \bibinfo{publisher}{Springer International Publishing},
  \bibinfo{address}{Cham}, \bibinfo{pages}{373--405}.
\newblock
\showISBNx{978-3-319-91590-6}
\showDOI{%
\url{https://doi.org/10.1007/978-3-319-91590-6_11}}


\end{thebibliography}

\end{document}